\pdfoutput=1
\documentclass{article}
\usepackage[utf8]{inputenc}

\newcommand{\sltr}{\mathrm{SL}(2,\mathbb{R})}

\newenvironment{eqaed}
    {
    \begin{equation}
        \begin{aligned}
    }
    {
        \end{aligned}
    \end{equation}\ignorespacesafterend
    }

\usepackage{amsmath,amsthm,amssymb}
\usepackage[dvipsnames]{xcolor}
\usepackage{physics}
\usepackage{tensor}
\usepackage{graphicx}
\usepackage{float}
\usepackage{xcolor, colortbl}
\usepackage{hyperref}
\usepackage{authblk}
\usepackage{mathtools}
\usepackage{geometry}
\definecolor{light-gray}{gray}{0.9}

\usepackage{setspace}
\usepackage{xparse}
\usepackage{esvect} 
\usepackage{slashed} 
\usepackage{dsfont} 
\usepackage[dvipsnames]{xcolor} 
\usepackage{empheq}
\usepackage{makecell}
\usepackage{subcaption}
\usepackage{float}
\usepackage{hyperref}
\hypersetup{
    colorlinks=true,
    linkcolor=blue,
    filecolor=magenta,      
    urlcolor=cyan,
}
\usepackage{import}
\usepackage{makeidx}
\title{\sc{Sequential Monte Carlo with Cross-validated Neural Networks for Complexity of Hyperbolic Black Hole Solutions in 4D }}

\author[1]{ Armin Hatefi\footnote{ahatefi@mun.ca }}
\author[2]{ Ehsan Hatefi\footnote{ehsan.hatefi@uah.es, ehsanhatefi@gmail.com}}
\affil[1]{Department of Mathematics and Statistics, Memorial University of Newfoundland, St John’s, NL, Canada.}
\affil[2]{University of Alcal\'a, Department of Signal Theory and Communications, Research group GRAM, Alcal\'a de Henares, Spain.}

\begin{document}

\maketitle

\vspace{-0.7cm}

\begin{abstract}

This paper investigates the self-similar solutions of the Einstein-axion-dilaton configuration from type IIB string theory and the global SL(2,R) symmetry. We consider the Continuous Self Similarity (CSS), where the scale transformation is controlled by an SL(2, R) boost or hyperbolic translation. The solutions stay invariant under the combination of space-time dilation with internal SL(2,R) transformations.
We develop a new formalism based on Sequential Monte Carlo (SMC) and artificial neural networks (NNs) to estimate the self-similar solutions to the equations of motion in the hyperbolic class in four dimensions. Due to the complex and highly nonlinear patterns, researchers typically have to use various constraints and numerical approximation methods to estimate the equations of motion; thus, they have to overlook the measurement errors in parameter estimation.
Through a Bayesian framework, we incorporate measurement errors into our models to find the solutions to the hyperbolic equations of motion. It is well known that the hyperbolic class suffers from multiple solutions where the critical collapse functions have overlap domains for these solutions. To deal with this complexity, for the first time in literature on the axion-dilaton system, we propose the SMC approach to obtain the multi-modal posterior distributions. Through a probabilistic perspective, we confirm the deterministic $\alpha$  and $\beta$ solutions available in the literature and determine all possible solutions that may occur due to measurement errors. We finally proposed the penalized Leave-One-Out Cross-validation (LOOCV) to combine the Bayesian NN-based estimates optimally. The approach enables us to determine the optimum weights while dealing with the co-linearity issue in the NN-based estimates and better predict the critical functions corresponding to multiple solutions of the equations of motion.

\vskip.12in

\end{abstract}

\newpage

\section{Introduction}\label{sec:intro}

It is well known that all Black holes can be characterised by their mass, their charge  as well as their angular momentum. Choptuik in \cite{Chop} also showed that there is yet one more parameter that explains the critical gravitational collapse solutions and it is called the critical exponent.
More specifically, Christodolou in \cite{Christodolou} first had revealed the spherically symmetric collapse of the real scalar field. Later on, Choptuik \cite{Chop} numerically showed that the real scalar field gravitational collapse solution demonstrates the discrete self-similarity property. Indeed, the gravitational solution shows space-time self-similarity where the dilations can take place. Therefore, the critical solution does provide an scaling law. If we show the initial condition of the real scalar field by parameter $p$, which is called the field amplitude, then $p=p_\text{crit}$ defines the critical solution and hence, the black hole can be  formed once $p$ chooses the values bigger than $p_\text{crit}$. Indeed, for $p>p_\text{crit}$  the mass of the black hole or the Schwarzschild radius are given by a scaling law as follows
\begin{equation}
r_S(p) \propto M_\text{bh}(p) \propto (p-p_\text{crit})^\gamma\,.
\end{equation}

 The following articles
 ~\cite{Chop,Hamade:1995ce,Hamade:1995ce24} found that the critical exponent for a real scalar field is given by  $\gamma\simeq 0.37$ in four dimensions. Notice that for dimensions bigger than four ($d \geq 4$), the black hole's mass scaled by \cite{KHA,AlvarezGaume:2006dw} as
\begin{equation}
 r_S(p) \propto (p-p_\text{crit})^\gamma \,, \quad M_\text{bh}(p) \sim (p-p_\text{crit})^{(D-3)\gamma}  \,.
\end{equation}

One may read some other numerical investigations for several other matter content in \cite{Birukou:2002kk,Husain:2002nk,Sorkin:2005vz,Bland:2005vu,HirschmannEardley,Rocha:2018lmv}. The collapse solutions of the perfect fluid had been studied in \cite{AlvarezGaume:2008qs,evanscoleman,KHA,MA} and its critical exponent $\gamma \simeq 0.36$ was also found in \cite{evanscoleman}.  The authors in \cite{Strominger:1993tt} argued that $\gamma$ might have a universal value for all matter fields that can be coupled to gravity in four dimensions. As discussed in ~\cite{KHA,MA,Hirschmann:1995qx} the critical exponent is explored by using the perturbations of self-similar solutions.

The solutions with axial symmetry were investigated in \cite{AE}, while shock waves are studied in \cite{AlvarezGaume:2008fx}. The authors in \cite{Hirschmann_1997}
explored the value of the critical exponent $\gamma \simeq 0.2641$ for the axion-dilaton configuration in four dimensions. 
Interestingly, in \cite{Antonelli:2019dqv} we have studied the perturbations and were able to precisely generate the existing value~ \cite{Hirschmann_1997} of $\gamma \sim 0.2641$ in four dimensions and other critical exponents have been derived in \cite{Hatefi:2020gis} in four and five dimensions.

Albeit the authors in \cite{Eardley:1995ns, Hamade:1995ce} investigated the entire analysis for the elliptic case in four dimensions, it is worth mentioning that their methods can also be examined in hyperbolic and parabolic cases as well as other dimensions, where for further results we refer to \cite{Hatefi:2020jdr, Hatefi:2020gis}.

Let us provide various motivation for the study of the critical collapse of the Einstein-axion-dilaton configuration. The first motivation is related to the gauge-gravity duality \cite{maldacena, wittenone, klebanov,wittentwo}, corresponding  the choptuik exponent, the imaginary part of quasi-normal modes as well as the dual conformal field theory that is pointed out in \cite{Birmingham:2001hc}. In fact, of the interest is to study the spaces that approach asymptotically to $AdS_5\times S^5$, where the simplest system is to consider in type IIB string theory of the axion-dilaton system with the self-dual 5-form field. Then, one can start to analyze the black hole solutions in diverse dimensions. It is also important to highlight that this system has also been related to the holographic description of black hole formation, see  \cite{scalingqcd,AlvarezGaume:2008qs}.  Finally the implications of this system to black hole physics  have been carried out in \cite{Hatefi:2012bp,Ghodsi_2010}. The key role of S-duality for these self similar solutions  has also been considered in \cite{Hamade:1995jx}. In this paper, we are dealing with the collapse of matter to form small mass black holes and hence one considers a small space-time region just close to where the singularity occurs. It has also been shown that this event is independent of the asymptotic structure of the space-time to which the collapse happens. In fact there is already the numerical evidence in asymptotically AdS space-times confirming that this is the case \cite{Husain:2002nk}. Therefore we eliminate the cosmological constant and just analyse self-similar solutions for the axion-dilaton system.

The self-similar solutions for all elliptic, hyperbolic and parabolic classes of SL(2,R) have been discovered in \cite{ours} in four and five dimensions for all classes of SL(2,R), which are the extensions of the earlier results \cite{AlvarezGaume:2011rk,hatefialvarez1307}.  In 
\cite{Hatefi:2021xwh} we have also recently made use of the Fourier-based regression models for obtaining the critical solutions. Consequently, the challenges of \cite{Hatefi:2021xwh} have been addressed in \cite{Hatefi:2022shc} where we applied truncated power basis, natural spline and penalized B-spline regression models  accordingly in order to be able to explore the non-linear functions. In \cite{Hatefi:2023vma} we applied artificial neural networks in order to address the instability of the black hole solutions for the specific parabolic class in higher dimensions. Lastly, in \cite{Hatefi:2023sgr}, we actually proposed a new formalism to be able to model the complexity of elliptic black hole solution in four dimension using hamiltonian monte carlo with stacked neural networks.

In this paper, we propose a new formalism based on Sequential Monte Carlo (SMC) and artificial neural networks (NNs) to be able to model the hyperbolic class of the spherical gravitational self-similar solutions in four dimensions. Due to the nature of highly non linear equations of motions of hyperbolic black holes, various authors used a variety of numerical calculations to simplify the equations of motions and parameters of the theory, for instance one can see \cite{Hatefi:2020jdr,Hatefi:2022shc,Hatefi:2023vma}. Hence, due to this reason the authors must have overlooked the measurement errors that are imposed
in exploring the parameters through various numerical methods.

Thus here we propose a new method to carry out the measurement errors, involved in parameter estimation, into our statistical models in exploring the solutions to the equations of motion. Recently Hatefi et al. \cite{Hatefi:2023sgr} applied the Hamiltonian Monte Carlo method to find solutions to the equations of motion in the elliptic class of four dimensions in a Bayesian framework. Unlike \cite{Hatefi:2023sgr}, the hyperbolic equations of motion in  four dimensions have  multiple solutions and therefore the collapse functions do have overlap domains under these solutions. In order to deal with this challenge, for the first time in the literature on the axion-dilaton system, we proposed the SMC approach to derive the posterior distribution of the parameters. The posterior distribution does provide all possible solutions in estimating the parameter of the equations of motion. Interestingly, the posterior distribution confirms the deterministic $\alpha$ and $\beta$ solutions found in the literature for the hyperbolic class in four dimensions. 
Unlike other methods in the literature, in this paper, we impose the $l_2$ penalized Leave-One-Out Cross-validation (LOOCV) to optimally combine the Bayesian NNs candidates.  The advantage of this approach is that it also enables us to determine the optimum weights while dealing with the co-linearity issue in the NN-based estimates and better predict the critical functions corresponding to multiple solutions of the equations of motion in the hyperbolic class.

The organization of the paper is as follows.  In section \ref{sec:problem} 
we briefly explain the relevant effective action for the axion-dilaton configuration, its equations of motion as well as the initial conditions that come from the  continuous self-similarity requirement. We the  describe our methodology to actually model the complexity of Sequential Monte Carlo with Cross-validated
Neural Networks for hyperbolic black hole solutions in  four dimensions. In section \ref{sec:numerical_studies}, we  use SMC samples from the posterior distribution and construct NN estimates based on the posterior mean and LOOCV as well as the 95\% credible intervals in estimating  the critical collapse functions corresponding to multiple solutions of the equations of motion.  Lastly, we present the results and conclude in the section \ref{sec:conclusions}.

\section{The Einstein-Axion-Dilaton System and Its Equations of Motion for Hyperbolic Class}\label{sec:problem}

The two real scalar fields of axion and dilaton can be combined to construct a single complex scalar field $ \tau \equiv a + i e^{- \phi}$.  Its dynamics and coupling to the gravity or its  effective action for four-dimensional axion-dilaton ($a,\Phi$) system is described by
\begin{equation}
S = \frac{1}{16 \pi G}\int d^4 x \sqrt{-g} \left( R - \frac{1}{2}  \frac{\partial_a \tau
\partial^a \bar{\tau}}{(\mathop{\rm Im}\tau)^2} \right),
\label{eaction}
\end{equation}
where $R$ is the scalar curvature. If we take variations from the metric and $ \tau$ then one  would be able to find out all the equations of motion as follows
\begin{equation}
\label{eq:efes}
R_{ab} = \frac{1}{4 (\Im\tau)^2} ( \partial_a \tau \partial_b
\bar{\tau} + \partial_a \bar{\tau} \partial_b \tau)\;,
\end{equation}
\begin{equation}\label{eq:taueom}
\nabla^a \nabla_a \tau +\frac{ i \nabla^a \tau \nabla_a \tau }{
\Im\tau} = 0 \,.
\end{equation}
 
 This effective action is classically invariant under SL(2,R) transformations which means that if $\tau$ gerts replaced by
\begin{equation}
\tau \rightarrow  \frac{a\tau+b}{c\tau+d},
\label{sltwo}
\end{equation}
where $(a,b,c,d) \in R$, $ad - bc = 1$ then $g_{ab}$ and the action remains invariant.  As argued originally by \cite{sen,greenschwarzwitten, polchinski,Font:1990gx}) this group gets broken to SL(2,Z) as an indication of duality transformation.

The spherically symmetric metric is represented by \cite{Eardley:1995ns} 
\begin{equation} 
	ds^2 = \left(1+u(t,r)\right)\left(- b(t,r)^2dt^2 + dr^2\right)
			+ r^2d\Omega^2_{d-2} \; .
\label{metric1}
\end{equation}

If we take into account the time scaling for (\ref{metric1}),  (as shown in \cite{Eardley:1995ns}), one can set $b(t,0)=1$ and regularity condition indicates  $u(t,0)=0$. The so called continuous self-similarity (CSS) means there exist a killing vector $\xi$ that generates global scale transformation,
where in spherical coordinates, we define $\xi=t\,\partial/\partial t+r\,\partial/\partial r$. 
The assumption of continuous scale invariance for the metric gets related a scaling for the line element under dilations as follows
\begin{equation}
    (t,r)\rightarrow ( \Lambda t,\Lambda r)\,,\quad \Lambda>0
\end{equation}
then
\begin{equation}
    ds^2 \rightarrow \Lambda^2 ds^2\, ,
\end{equation}
Now if we consider the scale invariant variable $z=-r/t$ then the self-similarity of the metric implies that the functions $u(t,r), b(t,r)$ must be expressed in terms of $z$, that is
\begin{equation}\label{eq:metricscaling}
    u(t,r) = u(z)\,,\quad b(t,r) = b(z)\,,
\end{equation}

The scalar $\tau$ must also be invariant up to an $\sltr$ transformation, so that
\begin{equation}\label{eq:tauscaling}
    \tau(t,r) \rightarrow M(\Lambda) \tau(t,r)\,.
\end{equation}
Hence a system of $(g,\tau)$ that satisfies eqts.~\eqref{eq:metricscaling},~\eqref{eq:tauscaling} to be  continuously self-similar (CSS). Hence, physically distinct cases are related to the different conjugacy classes of $\eval{\dv{M}{\Lambda}}_{\Lambda=1}$.

The effective action in \eqref{eaction} is SL(2,R)-invariant, hence we can consider a compensation of the scale transformation of $(t,r)$ by an SL(2,R) transformation. In fact in \cite{AlvarezGaume:2011rk} we already found out three different possible assumptions for this particular system. Those were called the elliptic, hyperbolic and parabolic classes that are related to three classes of $SL(2,R)$ transformations which were used to compensate for a scaling transformation in space-time. Let us describe the hyperbolic ans\"atze for $\tau(t,z)$.

The general form of the ansatz for the hyperbolic class is given by 
\begin{equation}  
\tau(t,r) = \frac{1-(-t)^w f(z)}{1+(-t)^w f(z)}, \label{tauansatz_hyperbolic}\end{equation}

where under a scaling transformation $t\rightarrow \lambda\, t$, $\tau(t,r)$ changes by a $SL(2,R)$ boost or hyperbolic translation, which means that all equations are invariant under the following transformation\begin{equation} f(z) \rightarrow e^\lambda f(z)\,,\quad \lambda \in \mathbb{R}.\end{equation}
Note that under $SL(2,R)$-transformation the following  \begin{equation} 
\tau(t,r)\rightarrow  (-t)^{\omega}\,f(z)\end{equation}

exactly produces the same equations of motion for hyperbolic case, where $f(z)$ is a complex function satisfying $\Im f(z)>0$, and $\omega$ is a real constant. 
Let us describe the derivation of the equations of motion for the hyperbolic class in four dimensions.  If we apply continuous self-similarity ans\"atze 
\eqref{tauansatz_hyperbolic} to all the equations of motion \eqref{eq:efes} and \eqref{eq:taueom} then one would be able to explore the ordinary differential equations for $u(z)$, $b(z)$, $f(z)$. However, if we make use of the spherical symmetry then one reveals that $u(z)$ and its first derivation $u'(z)$ can be expressed in all the equations in terms of $b(z)$, $f(z)$ and their first derivatives as follows
\begin{eqnarray}\label{eq:u0explicit}
    u(z)& = &  \frac{z b'(z)}{b(z)}\\
    \frac{u'(z)}{(1+u(z))}  & = & \frac{w \bar f(z) f'(z)+w  f(z) \bar f'(z) - 2z  \bar f'(z) f'(z)}{(f(z)-\bar f(z))^2} 
\end{eqnarray}

All the ordinary differential equations (ODEs) are given by 
\begin{align}\label{eq:unperturbedbp} b'(z) & = B(b(z),f(z),f'(z))\,, \\  f''(z) & = F(b(z),f(z),f'(z))\,. \label{eq:unperturbedfpp}\end{align}
Finally, the equations of motion in the hyperbolic class in four dimensions are represented by

\begin{eqnarray}
b' & = &  { z(b^2 - z^2) \over b (f -\bar f)^2} f' \bar{f}' - {
 \omega (b^2 - z^2) \over b (f -\bar f)^2} (f \bar{f}'+ \bar{f} f')
- {\omega^2 z |f|^2 \over b (f -\bar f)^2} \end{eqnarray}
\begin{eqnarray}
f''& = & 
     - {z (b^2 + z^2) \over b^2 (f -\bar f)^2} f'^2 \bar{f}'
     + {2 \over (f -\bar f)} \left(\frac{1}{\bar f} 
       + { \omega (b^2 + z^2) \over  2b^2 (f -\bar f)} \right) \bar{f} f'^2,\nonumber \\&&
     + { \omega (b^2 + 2 z^2) \over b^2 (f -\bar f)^2} f f'
\bar{f}' 
  + {2 \over z} \left(-1 + { \omega z^2 (f +\bar f) \over (b^2 - z^2)
(f -\bar f)}\right.\nonumber \\&& 
+ \left.{\omega^2 z^4 |f|^2 \over b^2 (b^2 - z^2) 
(f -\bar f)^2}\right) f'- {\omega^2 z \over b^2 (f -\bar f)^2} f^2
\bar{f}'  \nonumber \\&&
+{2 \omega \over (b^2 - z^2)} \left(-\frac{1}{2} - { \omega (f +\bar f )
\over 2(f -\bar f)}\right.
- \left.{\omega^2 z^2 |f|^2 \over 2b^2 (f -\bar f)^2}
\right) f.
\label{1fzeom321}
\end{eqnarray}

The equation of motion for $b$ is a first-order linear in-homogeneous equation with initial condition $b(0)=1$. The initial conditions for $f(z), f'(z)$ are also realised by applying the smoothness of the critical solution.  From \eqref{1fzeom321},  we encounter five singularities at $z = \pm 0$, $z = \infty$ and $z = z_\pm$. One can readily show that the singularities
$z = \pm 0$, $z = \infty$ can be eliminated by the coordinate transformation as shown in \cite{hatefialvarez1307}. The last two
singularities are demonstrated by  $b(z_\pm) = \pm z_\pm$. They correspond with the horizon where $z=z_+$ is just a coordinate singularity as shown in ~\cite{Hirschmann_1997,AlvarezGaume:2011rk}. Therefore by definition, $\tau$ must also be regular across it. 

\vskip.1in

Hence $f''(z)$ must also be finite as $z\rightarrow z_+$. Therefore, one finds that the vanishing of the divergent part of $f''(z)$ produces a complex-valued constraint at $z_+$, that is indicated by  $G(b(z_+), f(z_+), f'(z_+)) = 0$ where the explicit form of the $G$ function for the hyperbolic  case in four dimension is given in 

\begin{eqaed}
    G(f(z_+),f'(z_+)) = &\, \bar{f}(z_+) \left(2 z_+ \left(2 \omega^2\right) f'(z_+)+2 (\omega-1) \omega \bar{f}(z_+)\right)\\& +f(z_+) \left(2 z_+ (-2+2 \omega+2) f'(z_+)+2 \omega \bar{f}(z_+) \left(2-\omega^2\right)\right)\\& -\frac{2 z_+ \bar{f}(z_+)^2 (2+2 \omega-2) f'(z_+)}{f(z_+)}-2\, \omega (\omega+1) f(z_+)^2\,.
\end{eqaed}

One finds it convenient to make use of change of variables as  $f(z) = u(z) + i v(z)$. Indeed, using regularity at $z=0$ and some residual symmetries one gets 

\begin{eqaed}\label{bcs1}
b(0) = 1,  \quad\quad\quad\quad f(0) = 1+i x_0 \quad\quad  (x_0>0)
\end{eqaed}
 as well as
\begin{equation}
        f'(0) =u'(0)=v'(0)=0 
\end{equation}

These equations are invariant under a constant scaling $f\rightarrow \lambda\ f$, therefore one has the freedom to choose either the real value of $f(z)$ or its imaginary part  as one wishes at a particular value of $z$, so we would like to set $u(0) = 1$. If one require the regularity at the origin $z=0$, then one explores the following initial conditions for the hyperbolic class as

\begin{eqaed}\label{bcs}
b(0) = u(0)= 1,  \quad\quad\quad\quad   u'(0) =v'(0)=0 
\end{eqaed}

Therefore, the real and imaginary parts of $G$ must vanish which determines $\omega$,  where  $(x_0>0)$ and $x_0$ is a real parameter. The three discrete solutions in four dimensions were explored in \cite{Antonelli:2019dqv} where these solutions are found by integrating numerically the equations of motion. The solutions in hyperbolic class are identified by seeing the very rapidly decreasing $\Im f(0)$. No solutions are explored with $\Im f(0)>1$. The $\alpha$ and $\beta$ solutions are evidently explored. The $\alpha$ solution is given by
\begin{equation}
\omega=1.362, \quad\quad v(0)=0.708, \quad\quad z_+=1.440 \nonumber
\end{equation}
The  $\beta$ solution is 
\begin{equation}
\omega=1.003, \quad\quad  v(0)=0.0822, \quad\quad  z_+=3.29 \nonumber
\end{equation}

However for the $\gamma$ solution we notice that due to the fact that $\Im f(0)$ is so small also the $z_+$ root-finding gets effected with numerical noise ( $\omega=0.541, v(0)=0.0059, z_+=8.44 \nonumber$), and hence the quality is not perfect  because the $G \sim 10^{-7}$ comparing to $G \sim 10^{-13}-10^{-17}$  for the first two solutions, hence we concentrate on $\alpha$, $\beta$ solutions.

It is also important to highlight the fact that in \cite{Hatefi:2020gis} we have shown that the Choptuik exponent depends on the dimensions, matter content as well as the different branches of the unperturbed self  similar solutions. Thus, we argue that the conjecture about the universality of Choptuik exponent is not  satisfied. However, we also claim that there may exist some other universal behaviours that could have been hidden in combinations of critical exponents or there might be some other parameters of the given theory that have not been considered yet in  our understanding of current investigations.

\section{Statistical Methods}\label{sec:stat}
In this section, we investigate Bayesian solutions to the parameters of the hyperbolic equations of motion  using sequential Monte Carlo methodology. 
Suppose ${\bf x}(t)= (x_1(t),\ldots,x_H(t))$ represent  the solutions to
\begin{align}\label{de}
\frac{d}{d t} x_i(t) = g_i({\bf x}(t) | {\bf \theta})\;,
\end{align}
where the system encompasses $H$ differential equations (DEs) in which $t$ denotes the space-time, $\theta$  encodes the set of all unknown parameters of the system and $x_i(t)$ denotes the solution to $i$-th DE,    $i=1,\ldots, H$. Henceforth, we call ${\bf x}(t)$ the DE variables. 

As the equations of motion in black holes are highly nonlinear, typically there is no closed-form solution for the true trajectory of the DE variables. Therefore, researchers often apply a series of numerical methods and observe the trajectory of the DE subject to measurement errors. 
Let $y_{ij}$ denote the observed trajectory of the $i$-th DE variable  at $j=1,\dots,n_i$ space-time points. 
To take into account the uncertainty involved in the observed DE variables, let $y_{ij}$ follow a Gaussian distribution  with mean 
$x_i(t_{j}|{\bf \theta})$ and standard deviation $\sigma_i$ for $i=1,\ldots,H$. 
Let ${\bf\Omega}=({\bf \theta},{\bf \sigma})$ denote the set of all unknown parameters of the model where  ${\bf \sigma} =(\sigma_1,\ldots,\sigma_H)$. Combing all information from the observed DEs, the likelihood function of ${\bf\Omega}$ is given by 
\begin{align}\label{ll}
p({\bf\Omega}|{\bf y}) 
= \prod_{i=1}^{H} \prod_{j=1}^{n_i} (\frac{1}{\sigma_i^2})^{-1/2} 
\exp\left\{-\frac{(y_{ij}- x_i(t_{j}|{\bf \theta}))^2}{2\sigma_i^2}
\right\}.
\end{align}
In real-world scenarios, while the DE systems depend on unknown parameters, researchers typically have prior knowledge about the parameters of the DE model. 
To incorporate the prior information into our estimation, we propose a Bayesian framework and treat  the unknown parameters of the DE system as random variables.  
This Bayesian estimation approach enables us to find the statistical distribution of the unknown parameters based on  the set of observed data, given prior information about the unknown parameters. 
The statistical distribution of parameters given the observed data is henceforth called the posterior distribution of the parameters 
$\pi({\bf \Omega}|{\bf y})$  where ${\bf \Omega}$ denotes the set of all unknown parameters  of the DE system. The posterior 
distribution then enables us to make statistical inferences about the uncertainty in the estimation procedure and quantify the characteristics of the DE system. 

\subsection{Sequential Monte Carlo}\label{sub:smc}
When we face DE system \eqref{de}, it is reasonable to take into account the effect of space-time argument and the sequential process of the DE observations ${\bf y}_1,\ldots,{\bf y}_t$. This facilitates updating the posterior distribution of the unknown parameters sequentially as the DE variables are sequentially observed in the system.   
Sequential Monte Carlo (SMC), as a simulation-based approach, is a flexible technique in Bayesian statistics to sequentially estimate the posterior distribution.  
Due to the advent of cheap and powerful computing resources, the SMC appears a convenient tool to implement, in a parallel fashion,  sequential computation of the posterior distribution in a general setting. 

Let $\{{\bf \Omega}_t, t \in {\mathbb N} \}$ denote the Markov process showing the trajectory of unknown parameters over space-time with the prior distribution $\pi({\bf \Omega}_0)$. The Markov property of the process indicates that the probability of process at space-time $t$ only depends on the previous step of the state; that is
\[
p({\bf \Omega}_t| {\bf \Omega}_0,\ldots, {\bf \Omega}_{t-1}) = 
p({\bf \Omega}_t | {\bf \Omega}_{t-1}),
\]
where $p({\bf \Omega}_t | {\bf \Omega}_{t-1})$ denotes the transition probability from ${\bf \Omega}_{t-1}$ to ${\bf \Omega}_t$ in the parameter space. 
Let $\{{\bf y}_t;t \in {\mathbb N}\}$ denote the sequence of the observations from the DE system where they are conditionally independent  given the observed status of the parameter process with distribution $p({\bf y}_t | {\bf \Omega}_t)$ for $t \ge 1$.
For the sake of convenience in notations, let ${\bf \Omega}_{0:t}$  and ${\bf y}_{1:t}$ 
represent  the parameter  and DE observation sequences up to space-time $t$, respectively; That is, ${\bf \Omega}_{0:t} = ({\bf \Omega}_{0}, \ldots, {\bf \Omega}_{t})$ and ${\bf y}_{1:t}= ({\bf y}_{1},\ldots, {\bf y}_{t})$.

In this section, our focus is to employ the SMC properties to obtain recursively the posterior distribution  $\pi({\bf \Omega}_{0:t}|{\bf y}_{1:t})$ at any space-time $t$. 
In order to do that, at space-time $t$, one can apply the Bayes rule and write
\begin{align}\label{bayes_role}
    \pi({\bf \Omega}_{0:t}|{\bf y}_{1:t}) =
    \frac{p({\bf y}_{1:t}|{\bf \Omega}_{0:t}) \pi({\bf \Omega}_{0:t})}{\int p({\bf y}_{1:t}|{\bf \Omega}_{0:t}) \pi({\bf \Omega}_{0:t}) d{\bf \Omega}_{0:t}}.
\end{align}
From the joint distribution of ${\bf \Omega}_{0:t}$ and ${\bf y}_{1:t}$, one can also recursively  update the posterior distribution 
$\pi({\bf \Omega}_{0:t}|{\bf y}_{1:t})$ based on the posterior distribution of previous lags of the process by
\begin{align}\label{bayes_role_t1}
    \pi({\bf \Omega}_{0:t}|{\bf y}_{1:t}) =
    \pi({\bf \Omega}_{0:t-1}|{\bf y}_{1:t-1}) 
    \frac{p({\bf y}_{t}|{\bf \Omega}_{t}) \pi({\bf \Omega}_{t}|{\bf \Omega}_{t-1})}{p({\bf y}_{t}|{\bf y}_{1:t-1})}.
\end{align}
Accordingly, one can compute the posterior expectation of any characteristic of the DE system by
\begin{align}\label{Ef_post}
    \mathbb{E}_{\pi({\bf \Omega}_{0:t}|{\bf y}_{1:t})}
    \left[
    \mathbb{H}_t({\bf \Omega}_{0:t})
    \right] =
    \int \mathbb{H}_t({\bf \Omega}_{0:t}) \pi({\bf \Omega}_{0:t}|{\bf y}_{1:t}) d{\bf \Omega}_{0:t}
\end{align}
Due to the complex structure of the uncertainty and multidimensional integration of the marginal distribution in the DE system, there is no analytical form for the posterior distributions  \eqref{bayes_role} and \eqref{bayes_role_t1}; thus the conditional expectation \eqref{Ef_post} is not tractable too.

Importance sampling \cite{Rubinstein, Robert_casella}, as a practical solution to the intractability problem, uses an instrumental distribution to sample indirectly from the posterior distribution $\pi({\bf \Omega}_{0:t}|{\bf y}_{1:t})$. 
Let $q({\bf \Omega}_{0:t}|{\bf y}_{1:t})$ represent an instrumental distribution whose domain includes the domain of the target posterior distribution \eqref{bayes_role}. Accordingly, one can employ the importance sampling and rewrite \eqref{Ef_post} as
\begin{align}\label{Ef_post_ip}
    \mathbb{E}_{\pi({\bf \Omega}_{0:t}|{\bf y}_{1:t})}
    \left[
    \mathbb{H}_t({\bf \Omega}_{0:t})
    \right] =
    \frac{\int \mathbb{H}_t({\bf \Omega}_{0:t}) \lambda({\bf \Omega}_{0:t}) q({\bf \Omega}_{0:t}|{\bf y}_{1:t}) d{\bf \Omega}_{0:t}}{\int  \lambda({\bf \Omega}_{0:t}) q({\bf \Omega}_{0:t}|{\bf y}_{1:t}) d{\bf \Omega}_{0:t}},
\end{align}
where the non-normalized importance weights $\lambda({\bf \Omega}_{0:t})$ are given by
\begin{align}\label{lamba_weigt}
\lambda({\bf \Omega}_{0:t}) =
\frac{\pi({\bf \Omega}_{0:t}|{\bf y}_{1:t})}{q({\bf \Omega}_{0:t}|{\bf y}_{1:t})}.
\end{align}
Let $\{{\bf \Omega}_{0:t}^{(i)}\}; i=1,\ldots,n$ represent $n$ independent and identically distributed particles from the instrumental distribution $q({\bf \Omega}_{0:t}|{\bf y}_{1:t})$. Thus, using the importance sampling method, the Monte Carlo estimate of the quantity of interest $\mathbb{H}_t(\cdot)$ is given by
\begin{align}\label{Ef_post_ip_est}
    \widehat{
    \mathbb{H}_t({\bf \Omega}_{0:t})} =
    \frac{ \sum_{i=1}^{n} \mathbb{H}_t({\bf \Omega}_{0:t}^{(i)}) \lambda({\bf \Omega}_{0:t}^{(i)})}{\sum_{i=1}^{n}  \lambda({\bf \Omega}_{0:t}^{(i)})}= 
    \sum_{i=1}^{n} \mathbb{H}_t({\bf \Omega}_{0:t}^{(i)}) \Lambda({\bf \Omega}_{t}^{(i)})
\end{align}
where $\Lambda({\bf \Omega}_{0:t}^{(i)})$, as normalized importance weights, are obtained by
\begin{align}\label{non_weight}
\Lambda({\bf \Omega}_{t}^{(i)}) =
\frac{\lambda({\bf \Omega}_{0:t}^{(i)})}{\sum_{i=1}^{n} \lambda({\bf \Omega}_{0:t}^{(i)})}.
\end{align}
The importance sampling estimator \eqref{Ef_post_ip_est}, as a general framework of Monte Carlo, is convenient to implement. Despite this convenience,  the iterative structure of the technique is not adequate to 
sequentially  incorporate the new status of the process into estimating the 
$\pi({\bf \Omega}_{0:t}|{\bf y}_{1:t})$. As the new status of the sequence becomes available, one has to recompute all the important weights on the entire parameter space. 
This becomes computationally expensive for complex and nonlinear equations of motion.

Sequential Monte Carlo (SMC) \cite{Robert_casella, Moral}, as a sequential architecture of importance sampling, can be considered as a solution to the iterative problem of general importance sampling. 
 As observed from \eqref{lamba_weigt}-\eqref{non_weight}, in importance sampling, when a new status of the DE variable becomes available $y_t$, one has to re-compute the posterior distribution of the entire trajectory ${\bf \Omega}_{1:t}$ given the sequence ${\bf y}_{1:t}$. Hence, one requires re-computing even the importance weights of the previous states of the trajectory given the new status of the DE variable.
Unlike the importance sampling method,  the SMC sampler does not require re-computing the importance weights corresponding to the previous states of the trajectory $({\bf \Omega}_0,\ldots,{\bf \Omega}_{t-1})$, when the new status of the DE variable becomes available. In other words, the importance weights of the previous states of the trajectory stay the same, and we no longer need to re-compute them. We only need to compute the posterior distribution of the most recent state of the trajectory, given the new data.
 Treating the instrumental distribution  based on the previous states as the full marginal distribution, the instrumental distribution is thus updated by
\begin{align}\label{q_smc}
 q({\bf \Omega}_{0:t}|{\bf y}_{1:t}) = 
 q({\bf \Omega}_{0:t-1}|{\bf y}_{1:t-1}) 
 q({\bf \Omega}_{t}|{\bf \Omega}_{0:t-1},{\bf y}_{1:t}).
\end{align}
In a similar fashion as \eqref{q_smc},  one can  recursively show that 
\begin{align}\label{q_smc_omg0}
 q({\bf \Omega}_{0:t}|{\bf y}_{1:t}) = 
 \pi({\bf \Omega}_{0}) 
 \prod_{k=1}^{t} q({\bf \Omega}_{k}|{\bf \Omega}_{0:k-1},{\bf y}_{1:k}).
\end{align}
From the sequential representation \eqref{q_smc_omg0}, one can easily show that the non-normalized importance weights \eqref{non_weight} can be sequentially updated by
\begin{align}\label{smc_la}
\lambda({\bf \Omega}_{0:t}^{(i)}) =
\lambda({\bf \Omega}_{0:t-1}^{(i)})
\frac{p({\bf y}_{t}|{\bf \Omega}_{t}^{(i)}) \pi({\bf \Omega}_{t}^{(i)}|{\bf \Omega}_{0:t-1}^{(i)})}{ p({\bf y}_{t}|{\bf y}_{1:t-1}) \pi({\bf \Omega}_{t}^{(i)}|{\bf \Omega}_{0:t-1}^{(i)},{\bf y}_{1:t})},
\end{align}
and consequently, the normalized importance weights are sequentially updated by
\begin{align}\label{smc_LA}
\Lambda({\bf \Omega}_{t}^{(i)}) \propto
\Lambda({\bf \Omega}_{t-1}^{(i)})
\frac{p({\bf y}_{t}|{\bf \Omega}_{t}^{(i)}) \pi({\bf \Omega}_{t}^{(i)}|{\bf \Omega}_{0:t-1}^{(i)})}{\pi({\bf \Omega}_{t}^{(i)}|{\bf \Omega}_{0:t-1}^{(i)},{\bf y}_{1:t})}.
\end{align}
As a special case of \eqref{q_smc_omg0}, one can  employ the prior distribution in the SMC framework \cite{Robert_casella, Moral}. In this case, the instrumental distribution is given by
\begin{align}\label{q_smc_prior}
 q({\bf \Omega}_{0:t}|{\bf y}_{1:t}) = 
 \pi({\bf \Omega}_{0}) 
 \prod_{k=1}^{t} \pi({\bf \Omega}_{k}|{\bf \Omega}_{k-1}).
\end{align}
In this case, from \eqref{q_smc_prior} and the fact that 
$p({\bf y}_{t}|{\bf y}_{1:t-1})$ is constant over the entire trajectory of ${\bf \Omega}_{0:t}$, one can easily update the importance weights of the $i$-th particle by
\begin{align}\label{smc_LA_prior}
\Lambda({\bf \Omega}_{t}^{(i)}) \propto
\Lambda({\bf \Omega}_{t-1}^{(i)}) p({\bf y}_{t}|{\bf \Omega}_{t}^{(i)}).
\end{align}
Although the SMC technique is well suited to accommodate sequentially the new data into estimation, the posterior distribution of the particles  very quickly becomes skewed after only a few steps such that  only a few particles will have a non-zero probability \cite{Moral,Bernardo}. 
Consequently, the Monte Carlo chain will not be able to sample from all  aspects of target $\pi({\bf \Omega}_{0:t}|{\bf y}_{1:t})$. 
To deal with this degeneracy issue, a re-sampling is added to SMC to eliminate sequentially particles with low importance weights and at the same time augment the particles from high-density areas.
To implement the re-sampling step, one can take $n$ random draws with replacement from the collection of the particles $\{{\bf \Omega}_{0:t}^{(1)},\ldots, {\bf \Omega}_{0:t}^{(n)}\}$ with probabilities corresponding to the important weights $\{\Lambda({\bf \Omega}_{t}^{(1)}),\ldots,\Lambda({\bf \Omega}_{t}^{(n)})\}$.
Let $n_t^{(i)}$ denote the number of offsprings from the particle ${\bf \Omega}_{0:t}^{(i)}, i=1,\ldots,n$. It is easy to see that the particle survives and contributes to the posterior distribution when $n_t^{(i)} >0$; otherwise, the particle dies. The important sampling  and the re-sampling steps are finally alternated to update the important weights and filter the particles sequentially to obtain the samples from the posterior distribution. 

\subsection{Cross-validated Neural Networks} \label{sub:nns}
Neural networks (NNs), inspired by the human neural system, comprise a network of connected neurons. These connections enable the neurons to send information from one layer to another. 
According to the flexibility and power of the NNs, they have been increasingly exploited in recent years as a reliable predictive model for solving differential equations. The power of the NNs enables us to reformulate finding solutions to the DE system to a parametric estimation of the DE variables by minimizing the prediction errors. 
NNs consist of a multi-layer perceptron whose layers include neurons connecting the layers of the network to each other.  
These connections enable the NN to estimate the functional form of the DE variables. 
In this subsection, we describe how NNs use the parameter estimates developed by the SMC method, from Subsection \ref{sub:smc}, to predict the complex and nonlinear forms of the DE variables in the hyperbolic class of 4d.  

Let $\mathcal{N}({\bf x}(t|{\bf\Omega}),t,{\bf \phi})$ denote the NN estimate, consisting of $L$ hidden layers, for the DE variable  ${\bf x}(t)$ form DE system \eqref{de}.
Each neuron of the NN is connected with another in the next layer via a linear regression model
\begin{align}
    \left\{\begin{array}{lc}
    {\bf Z}_j^l = {\bf W}^l \text{NN}^{l-1}({\bf x}(t),t,{\bf \phi})+ {\bf b}^l, & \\
    \text{NN}^{l}({\bf x}(t),t,{\bf \phi})= a({\bf Z}_j^l),  & l=1,\ldots,L.    
    \end{array}
    \right.
\end{align}
such that $\text{NN}^{l-1}(\cdot)$ represents the response observed from the $l$-th layer, ${\bf \phi}=({\bf W}^1,\ldots,{\bf W}^L,{\bf b}^1,\ldots,{\bf b}^L)$ represents the set of all unknown parameters, $a$ denotes  a non-linear activation function, ${\bf W}^l$ and ${\bf b}^l$ show the weight matrix and bias vector of the $l$-th layer, respectively \cite{Bishop,Goodfellow:Deep}. 
Finally, the NN estimate of the DE variables ${\bf x}(t)$ are obtained as a solution to the squared loss function
\begin{align}\label{nn_loss}
   \widehat{\mathcal{N}}({\bf x}(t|{\bf\Omega}),t,{\bf \phi}) = \arg\min_{\bf\Omega} \left( \mathcal{N}({\bf x}(t|{\bf\Omega}),t,{\bf \phi}) - {\bf x}(t) \right)^2.
\end{align}
To handle the optimization \eqref{nn_loss}, the NNs implement a series of forward and backward propagation steps to find  the final solution to the DE system \eqref{de}. For more details about the theory and applications of the NNs, readers are referred to  \cite{Bishop, Lagaris, Goodfellow:Deep} and references therein.  

In this paper, we first plan to find the Bayesian estimate for the parameters of the equations of motion. The Bayesian proposals are then stacked into the equations of motion to find the Bayesian stacked NN solvers. 
To this end, as described in Subsection \ref{sub:smc}, we use the SMC  method and find the posterior distribution $\pi({\bf \Omega}|{\bf y})$. 
Let ${\bf \Omega}^*_{1},\ldots,{\bf \Omega}^*_{M}$ denotes $M$ Bayesian SMC proposals for the parameters of the DE system. 
Given the posterior candidate ${\bf \Omega}^*_{m}$, let $\widehat{\mathcal{N}}({\bf x}(t|{\bf \Omega}^*_{m}),t,{\bf \phi})$ denote the NN-based estimate of the DE variable ${\bf x}(t)$ for $m=1,\ldots,M$. From a probabilistic perspective, 
 the estimate $\widehat{\mathcal{N}}({\bf x}(t|{\bf \Omega}^*_{m}),t,{\bf \phi})$ will be the true solution to the DE system \eqref{de}  
with probability $\pi({\bf \Omega}^*_{m}|{\bf y})$ for $m=1,\ldots, M$.

Recently, Hatefi et al \cite{Hatefi:2023sgr} proposed the Bayesian model averaging to stack the NN-based estimates in predicting the critical solution for the elliptic class of 4d. Following \cite{Hatefi:2023sgr}, using the SMC candidates  ${\bf \Omega}^*_{1},\ldots,{\bf \Omega}^*_{M}$ and the training set ${\bf y}$ of size $n$, the posterior distribution of the $\widehat{\mathcal{N}}({\bf x}(t|{\bf \Omega}^*_{m}),t,{\bf \phi})$ at a fixed space-time $t$, is given by
\begin{align}\label{pots_nn}
    \mathbb{P}(\widehat{\mathcal{N}}({\bf x}(t|{\bf \Omega}),t,{\bf \phi})|{\bf y}) = \sum_{m=1}^{M} \mathbb{P}(\widehat{\mathcal{N}}({\bf x}(t|{\bf \Omega}^*_m),t,{\bf \phi})|{\bf y}) \pi({\bf \Omega}^*_m|{\bf y}). 
\end{align}
Using \eqref{pots_nn}, Hatefi et al \cite{Hatefi:2023sgr} proposed the mean posterior of the NN-based estimates to stack the NN candidates in predicting the DE variables in the elliptic class of equations of motion. 
Despite the simplicity of the posterior mean, when the posterior distribution has multiple high-density areas, the posterior mean may not be able to  capture the different solutions of the DE system. To handle the problem, we propose the idea of Leave-One-Out Cross-validation (LOOCV) to combine information from all high-density areas of the posterior and develop cross-validated NN-based estimates to more accurately estimate the DE variables under various solutions of the hyperbolic class equations of motion.   

Let ${\widehat{\mathcal{N}}({\bf x}|{\bf \Omega}^*)}=\left[\widehat{\mathcal{N}}_1({\bf x}(t|{\bf \Omega}^*_1),t,{\bf \phi}),\ldots,\widehat{\mathcal{N}}_M({\bf x}(t|{\bf \Omega}^*_M),t,{\bf \phi})\right]^\top$
 represent the design matrix of the $M$ estimates, corresponding to SMC candidates. 
 As $\widehat{\mathcal{N}}_1({\bf x}(t|{\bf \Omega}^*_1),t,{\bf \phi}),\ldots,\widehat{\mathcal{N}}_M({\bf x}(t|{\bf \Omega}^*_M),t,{\bf \phi})$ are $M$ estimates of the DE variable ${\bf x}(t)$ at space-time $t$, the estimates may be linearly dependent. This arises the co-linearity problem in the NN-based design matrix $\widehat{\mathcal{N}}({\bf x}|{\bf \Omega}^*)$. To this end, 
 under squared error loss with $l_2$ penalty, we can stack the NN-based estimates using the penalized linear regression model \cite{Bishop}. One can estimate the coefficients of the regression model ${\bf\beta} = (\beta_1,\ldots,\beta_M)$ by
 \begin{align}\label{ridge_reg}
     {\widehat\beta} = \arg\min_\beta \left[ \left( {\bf y} - \widehat{\mathcal{N}}({\bf x}|{\bf \Omega}^*)\right)^\top  \left( {\bf y} - \widehat{\mathcal{N}}({\bf x}|{\bf \Omega}^*)\right) + \lambda {\bf\beta}^\top {\bf\beta} \right].
 \end{align}
According to the properties of the least square under $l_2$ penalty, one can easily show that the solution to \eqref{ridge_reg} is given by
\begin{align}\label{ridge_est}
     {\widehat\beta} = \left[ \widehat{\mathcal{N}}({\bf x}|{\bf \Omega}^*)^\top \widehat{\mathcal{N}}({\bf x}|{\bf \Omega}^*)  + \lambda \mathbb{I}\right]^{-1} 
     \widehat{\mathcal{N}}({\bf x}|{\bf \Omega}^*) {\bf y}
 \end{align} 
When the  posterior distribution has multiple modes, the least square estimate \eqref{ridge_est} may assign unfair weights to some complex NN candidates based on the training set. To deal with this problem, we develop the LOOCV estimate of the weights where leaving one observation out in each iterative training step to find the best coefficient estimates.   Let $\widehat{\mathcal{N}}_m^{-i}({\bf x}(t|{\bf \Omega}^*_m),t,{\bf \phi})$ denote the NN-based estimate for the DE variables at space-time $t$ using the $m$-th SMC proposal when the $i$-th observation in the training set has been removed. 
From \eqref{ridge_reg}, the  LOOCV estimate of the weights are given by
\begin{align}\label{ridge_reg_loocv}
     {\widehat\beta}_{loocv} = \arg\min_\beta \left[ \sum_{i=1}^{M} \left( {\bf y} - \widehat{\mathcal{N}}^{-i}({\bf x}|{\bf \Omega}^*)\right)^\top  \left( {\bf y} - \widehat{\mathcal{N}}^{-i}({\bf x}|{\bf \Omega}^*)\right) + \lambda {\bf\beta}^\top {\bf\beta} \right].
 \end{align}
From \eqref{ridge_reg_loocv}, the LOOCV-based NN estimate of the DE variables at space-time $t$, using the SMC estimates ${\bf\Omega}^*_1,\ldots,{\bf\Omega}^*_1$ is given by 
\begin{align}\label{loocv_est}
     {\widehat{\bf x}(t)} = \sum_{m=1}^{M} {\widehat\beta}_{loocv} \,
     \widehat{\mathcal{N}}_m({\bf x}(t|{\bf \Omega}^*_m),t,{\bf \phi})
 \end{align}
The LOOCV-based NN estimates \eqref{loocv_est} cross-validates iteratively the NN candidates on $\widehat{\mathcal{N}}^{-i}({\bf x}|{\bf \Omega}^*)$; thus the final estimates avoid giving unfair weights to spurious SMC proposals for the training set.  It takes better into account the multiple high-density areas of the posterior distribution and consequently, it is expected to more efficiently estimate the DE variables of the system \eqref{de}, giving the SMC estimates ${\bf\Omega}^*_1,\ldots,{\bf\Omega}^*_1$.


\section{Numerical Studies}\label{sec:numerical_studies}
The equations of motion for hyperbolic class in four dimensions have five singular points; however, the relevant range for $z$, that contains the proper information lies between the two singularities
\begin{equation}
    z = 0\,,
\end{equation}
\begin{equation}
    z = z_+\,,\quad b(z_+) = z_+\,.
\end{equation}
In particular, $z=z_+$ is an event horizon, which is the homothetic horizon. Thus, it is a coordinate singularity, and $\tau$ must be regular across it, which is equivalent to the  finiteness of $f''(z)$ as $z\rightarrow z_+$. In fact the vanishing of the divergent part of $f''(z)$ leads to a complex-valued constraint at $z_+$ as 
\begin{equation}\label{eq:unpconstraint}
    \mathbb{C} \,\ni\, G(b(z_+), f(z_+), f'(z_+))\,.
\end{equation}
From time scaling, the regularity of $\tau$, and the residual symmetries in the equations of motions, one can show the initial boundary conditions as
\begin{equation}
        b(0) = 1, f(0)= 1+i x_0 , x_0>0,   \quad f'(0) =0\,, 
\end{equation}        
where $x_0$ is a real parameter as the initial value of the equations of motion. Thus, the system results in one parameter $\omega$ and two constraints which are the real and imaginary parts of $G$. Therefore, 
the system is handled by discrete solutions where the CSS solutions are numerically investigated.

Here we plan to estimate the self-similar solutions  using the Bayesian framework for the hyperbolic class in 4d. In this framework, we  treat the parameter of the equations of motion $\omega$ as a random variable and then use the SMC approach to find the posterior distribution.  The posterior distribution allows us to take into account the numerical measurement errors in estimating the critical collapse functions of the the system.

 The equations of motion of the axion-dilaton system have already been studied in various dimensions and also for different ansatz \cite{Hatefi:2023vma,Antonelli:2019dqv}. Hatefi et al. \cite{Hatefi:2022shc, Hatefi:2021xwh} applied  the statistical regresion models using Fourier-based and spline smothers to estimate the critical collapse functions. In a recent publication, Hatefi et al. \cite{Hatefi:2023vma} constructed a solver based on NNs and showed
  there was no solution in higher dimensions for the parabolic class of black holes. Particularly \cite{Antonelli:2019dqv} employed a root-finding method to numerically determine all the parameters of the equations of motion.

Due to the highly nonlinear equations of motion, researchers typically have to use a series of numerical approaches to simplify the equations and also to keep track of the parameters of the models. For instance, these techniques include imposing non-trivial constraints on the equations of motions such as finiteness of $f''(z)$ as $z\rightarrow z_+$. One might also point out one more constraint, namely,  the vanishing of the divergent part of $f''(z)$ that generates the complex-valued constraint at $z_+$. This implies that the real and imaginary parts of $G$ must vanish at $z\rightarrow z_+$. The equations in the hyperbolic case are not solvable analytically, hence \cite{ours} used the profile-root finding method and applied the discrete optimization on the coordinates of the extended parameter space of the equations. 
More specifically,  the equations are approximated by the first two orders of the Taylor expansions to make the root-finding step for the parameters of the equations tractable.
Next,  \cite{ours} could clearly estimate the parameters of the equations of motion by applying a grid search optimization on the coordinates  in the system.  After discarding the spurious roots in the domain, \cite{ours}  found out the solution to the equations of motion and then could estimate the critical collapse functions.

Although \cite{Hatefi:2023vma,Hatefi:2022shc, Hatefi:2021xwh, Antonelli:2019dqv} explored the hyperbolic class of equations motion  in deterministic approach, in this research  
through a stochastic perspective, we construct a Bayesian method using SMC approach to assess the measurement errors into the estimation of the critical functions. 
Recently Hatefi et al. \cite{Hatefi:2023sgr} proposed Bayesian  mean using Hamiltonian Monte Carlo to find solutions to the equations of motion in the elliptic class of 4d. Unlike \cite{Hatefi:2023sgr}, the hyperbolic equations of motion in 4d result in multiple solutions where the critical collapse functions have overlap domains under the multiple solutions. 
To deal with the complexity of the measurement errors in the hyperbolic equations of motion,  for the first time in the literature on the axion-dilaton system, we proposed the Sequential Monte Carlo approach to derive the posterior distributions of the parameters.
Unlike \cite{Hatefi:2023sgr} where they proposed Bayesian model averaging to stack the NN-based estimates, here we propose  the $l_2$ penalized Leave-One-Out cross-validation to stack the NN-based estimates for critical collapse functions. The approach enables us to assign the optimum weights to NN-based estimates and hence better estimate the critical functions corresponding to multiple solutions of the equations of motion.

The SMC Bayesian estimates provide information for all possible parameters' outcomes, that may take place in the numerical experiments, from the posterior distribution. They also allow researchers to embed this complexity  in estimating critical functions.
In this numerical study, we consider equations of motion \eqref{1fzeom321} as the DE system of interest. The system leads to three critical collapse functions $b_0(z), |f(z)|$ and $\arg(f(z))$ where we treated these functions as the DE variables  of the system that must be estimated.  Since all the DE variables of  system \eqref{1fzeom321} are numerically and simultaneously solved,  
it makes  sense to assume that the observed DE variables also have the same standard deviation $\sigma$ parameter which actually represents the variability in the numerical experiments. Hence, ${\bf \Omega}=(\omega,\sigma)$ encodes the set of all unknown parameters of the model. 
We used Python Package  Pymc3 \cite{pymc3} to implement the SMC approach and find the posterior distribution of ${\bf \Omega}$.
In order to investigate the effect of the prior information on the likelihood parameters in predicting of our critical functions, we assigned two different prior distributions for $\omega$. 
The prior distributions include non-informative uniform distribution between [0.3, 1.5]. We assign 
the second prior distribution to be the Gaussian distribution with a mean of $1.20$ and a standard deviation of $0.2$. We also consider the half-Cauchy distribution with scale parameter $0.5$ as the prior distribution for parameter $\sigma$ to capture the uncertainty involved in  the likelihood function. 

\begin{figure}[H]
    \centering
    \includegraphics[scale=0.5]{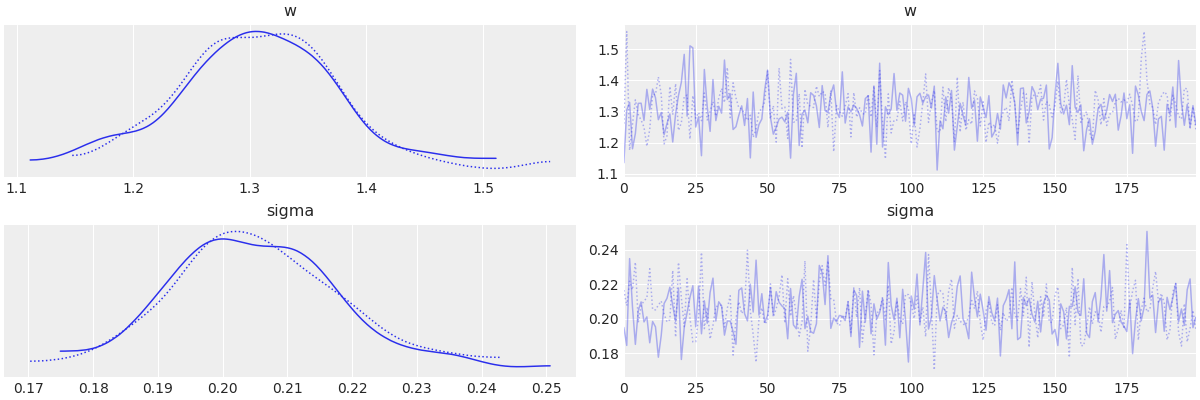}
    \caption{The posterior distributions and their trace-plots of two SMC chains (shown by solid and dotted blue lines) for parameters $\omega$ (w) and $\sigma$ (sigma) under Gaussian prior distribution.}
    \label{fig:nor_w}
\end{figure}
\begin{figure}[H]
    \centering
    \includegraphics[scale=0.5]{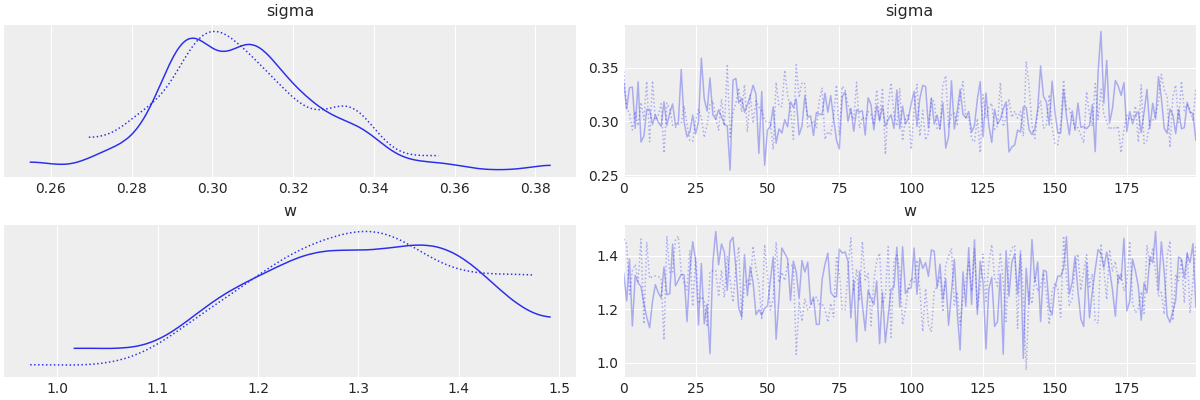}
    \caption{The posterior distributions and their trace-plots of two SMC chains (shown by solid and dotted blue lines) for parameters $\omega$ (w) and $\sigma$ (sigma) under Uniform prior distribution.}
    \label{fig:unif_w}
\end{figure}

We show the posterior distributions of parameters $\omega$ and $\sigma$ 
in Figures \ref{fig:nor_w} and \ref{fig:unif_w} under Gaussian and Uniform prior distributions, respectively. 
In each figure, we show two independent realizations of the SMC chain for the posterior distributions and their corresponding trace plots.  It is clear  that the posterior distributions of $\omega$ are multi-modal and have multiple high-density areas. Interestingly, this finding is compatible with the literature where \cite{ours} shows there must be at least three solutions to  hyperbolic equations of motion in 4d.
Under both prior distributions, we observe that there is a dominant high-density area almost ranging between [1.25, 1.4]. This corresponds to the 
$\alpha$-solution of \cite{ours}. In addition, we also observe that both figures confirm there is a small high density on the left tail of the posterior distribution of $\omega$ which is interestingly compatible with the $\beta$-solution to the equations of motion in \cite{ours}.
It should be noted that for example under Uniform prior, the probability that the deterministic $\alpha$-solution of \cite{ours} be the  true solution  to the hyperbolic equations motion is almost 10\%; that is 
$\pi(\omega \in 1.36 \pm 10^{-2}) \propto 0.10$. Moreover, the probability that the true solution appears to be smaller than 1.10 is given by $\pi(\omega \le 1.10) \propto 0.05$.

\begin{figure}[H]
    \centering
    \includegraphics[scale=0.36]{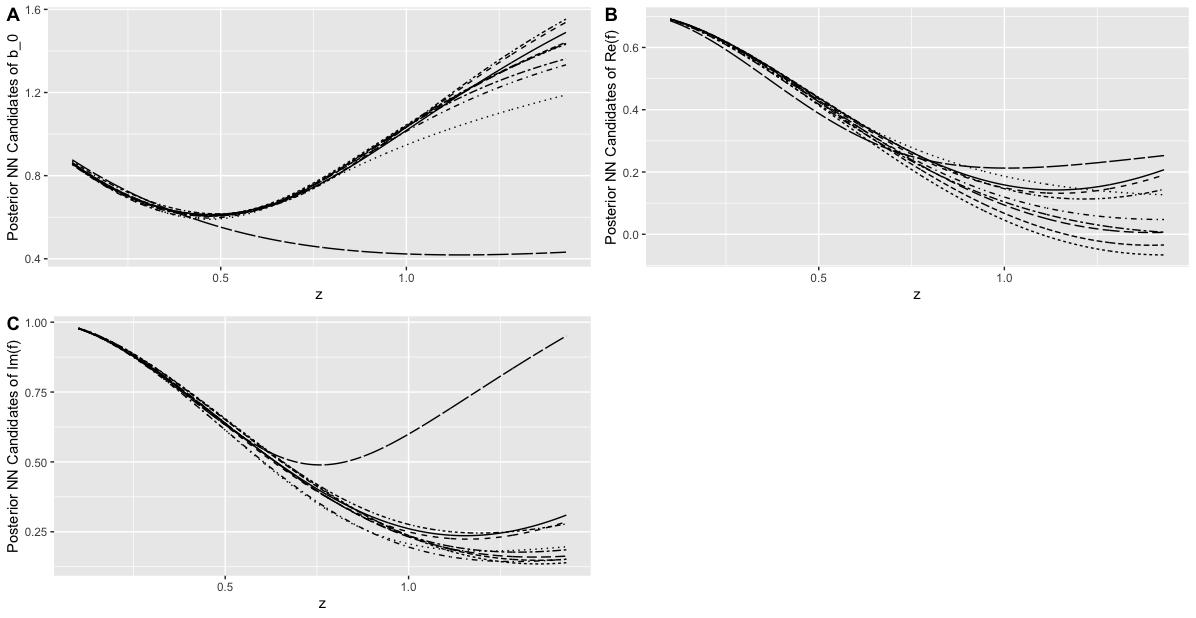}
    \caption{Ten randomly selected NN-based estimates for DE variables
    $b_0(z)$ ({\bf A}), $\text{Re}(f(z))$ ({\bf B}) and $\text{Im}(f(z))$ ({\bf C}) corresponding to ten SMC samples from the support of the posterior distribution $\pi(\omega|{\bf y})$ where Gaussian distribution was used as prior distribution. We show each estimate with a different line type.}
    \label{fig:nor_candids}
\end{figure}

\begin{figure}[H]
    \centering
    \includegraphics[scale=0.36]{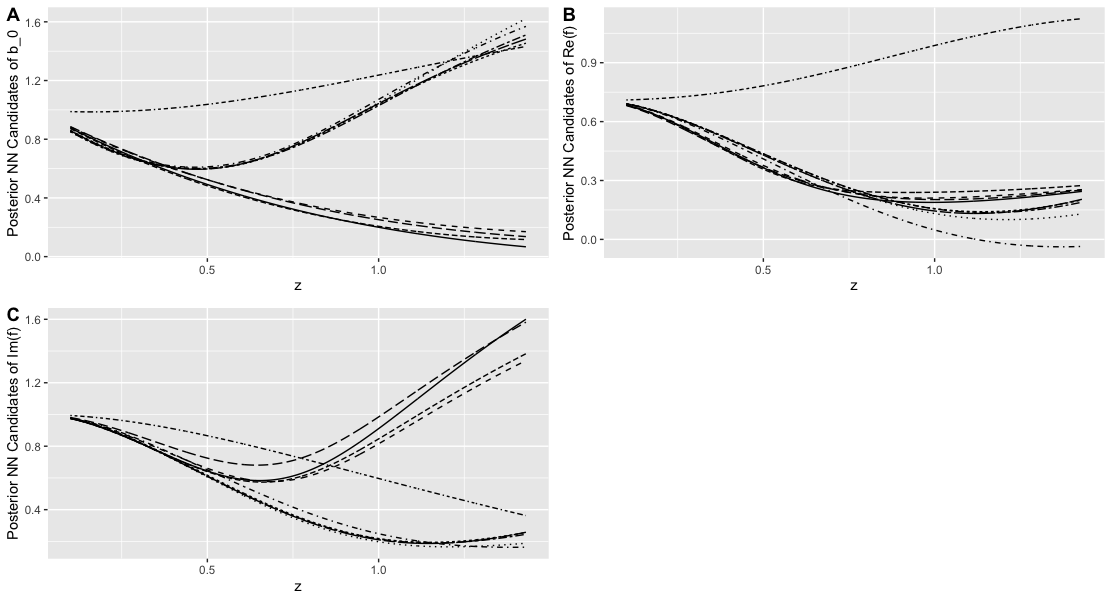}
    \caption{Ten randomly selected NN-based estimates for DE variables
    $b_0(z)$ ({\bf A}), $\text{Re}(f(z))$ ({\bf B}) and $\text{Im}(f(z))$ ({\bf C}) corresponding to ten SMC samples from the support of the posterior distribution $\pi(\omega|{\bf y})$ where Uniform distribution was used as prior distribution. We show each estimate with a different line type.}
    \label{fig:unif_candids}
\end{figure}

As described in Subsection \ref{sub:nns}, we construct the NN-based estimates of the critical collapse functions using the SMC proposals from the posterior distribution $\pi(\omega|{\bf y})$. 
To do that, we take $L=200$ samples ${\omega}^*_1,\ldots,{\omega}^*_L$ from $\pi(\omega|{\bf y})$.
Treating the posterior candidates ${\omega}^*_l,l=1,\ldots, L$ as the true value of the parameter in the equations of motion, we applied fully connected NNs to solve the DE system and find the NN-based estimates of the critical collapse functions. 
To do so, we used Python Package NeuroDiffEq \cite{Chen:2020} to carry out the neural networks for differential equations with 4 hidden layers where each layer consists of 16 neurons. We then ran the NNs for 1000 epochs and estimates the critical collapse function at 1000 equally spaced space-time points $z_i \in [0,1.44]$ for  $i=1,\ldots,1000$.
We finally obtained $L=200$ NN-based estimates for the critical collapse functions corresponding to $L$ realizations from the posterior distributions. 
From a probabilistic perspective, each of these $L$ realizations of NN-based estimates can occur to be the true form of the critical functions with probabilities $\pi({\omega}^*_l|{\bf y})$ for $l=1,\ldots, L$.
To better represent this probabilistic perspective and sampling variability from $\pi({\omega}^*_l|{\bf y})$, we show ten randomly selected realizations of the posterior NN-based estimates for the critical function  in Figures \ref{fig:nor_candids} and \ref{fig:unif_candids} under Gaussian and Uniform priors. We see that  two different patterns are almost observed for critical collapse functions on the same domain which is compatible with the multiple solutions available in the literature for the equations of motion under the hyperbolic class of 4d. 
In the next step of the numerical study, we plan to stack the $L=200$ NN-based candidates in estimating the critical collapse functions.  
Using a root-finding based on general relativity, \cite{ours} showed that there are three solutions in the domain of the equations of motion in the hyperbolic class of 4d. These solutions correspond to $\omega_1=1.362$ ($\alpha$-solution), $\omega_2=1.003$ ($\beta$-solution) and $\omega_3=0.005$ ($\gamma$-solution). It was discussed in  \cite{ours} that the $\gamma$-solution is not a stable solution due to the fact that $\Im f(0)$ is so small also the $z_+$ root-finding gets affected by numerical noise ( $\omega=0.541, v(0)=0.0059, z_+=8.44 \nonumber$), and hence the quality is not perfect and so it may be a spurious solution due to numerical noise. For this reason, in this research, we focused on the two $\alpha$ and $\beta$ solutions as two available solutions to the equations of motion in the literature.  
The critical collapse functions corresponding to $\alpha$-solution range between [0, 1.44], while the functions range between [0, 3.29] corresponding to $\beta$-solution. In order to investigate the performance of developed models in estimating the functional form of the  critical collapse  functions under both  solutions, we focus on the common areas between the two scenarios and ran investigated the NN-based estimates in [0,1.44].

\begin{figure}[H]
    \centering
    \includegraphics[scale=0.42]{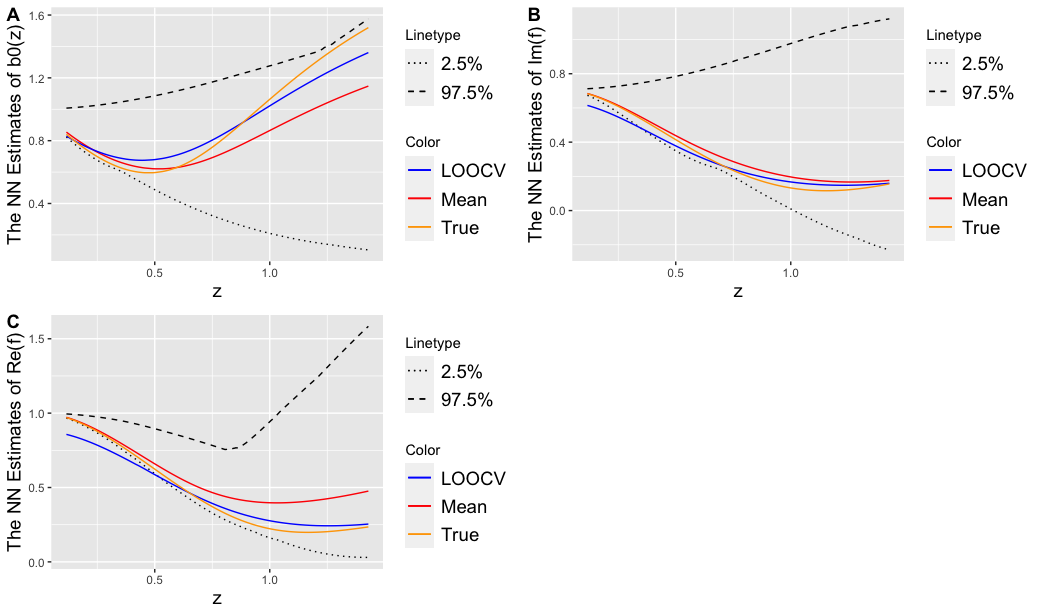}
    \caption{The posterior mean (red), LOOCV (blue) and the true, using $\omega=1.362$, (orange) NN-based estimates of the critical collapse functions $b_0(z)$ ({\bf A}), $\text{Im}(f(z))$ ({\bf B}) and  $\text{Re}(f(z))$ ({\bf C}) corresponding to the $\alpha$-solution scenario of the hyperbolic equations of motion. The dotted and dashed  lines show, respectively, the lower and upper bounds of the 95\% Bayesian credible intervals under Gaussian prior.  }
    \label{fig:nor_alpha_3_1}
\end{figure}
\begin{figure}[H]
    \centering
    \includegraphics[scale=0.39]{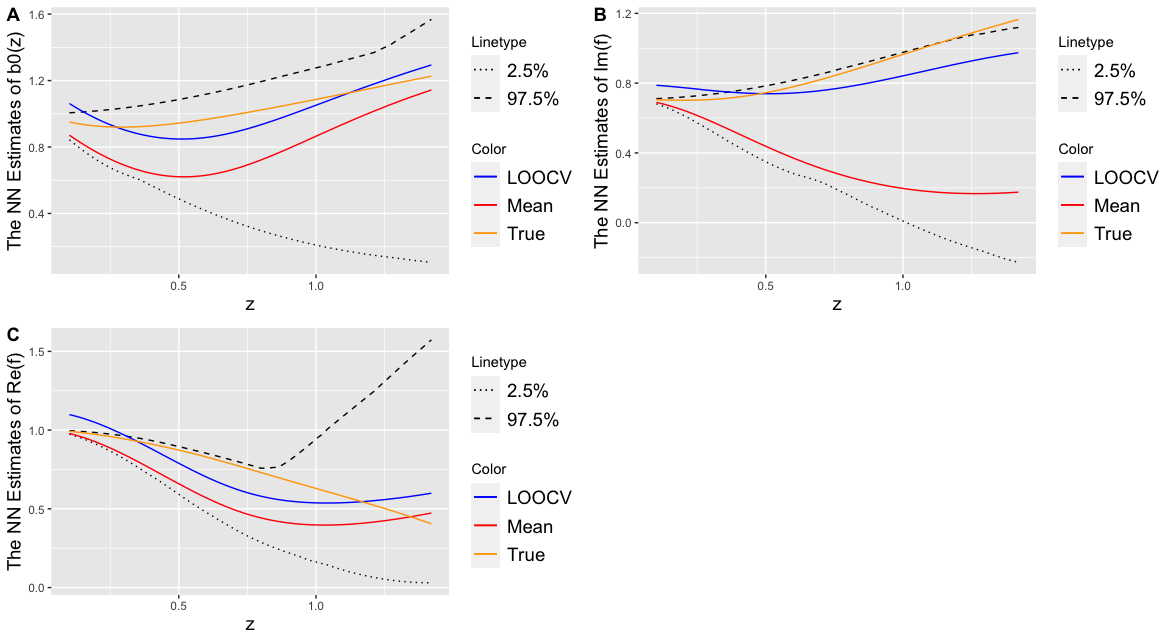}
    \caption{The posterior mean (red), LOOCV (blue) and the true, using $\omega=1.003$, (orange) NN-based estimates of the critical collapse functions $b_0(z)$ ({\bf A}), $\text{Im}(f(z))$ ({\bf B}) and  $\text{Re}(f(z))$ ({\bf C}) corresponding to the $\beta$-solution scenario of the hyperbolic equations of motion. The dotted and dashed  lines show, respectively, the lower and upper bounds of the 95\% Bayesian credible intervals under Gaussian prior.}
    \label{fig:nor_cv_beta_3_1}
\end{figure}
Following Hatefi et al \cite{Hatefi:2023sgr}, we first applied the Bayesian model averaging method and computed the posterior mean of the $L$ NN-based estimates at each space-time point $z_i,i=1,\ldots,1000$. Henceforth this estimate is called the mean stacked NN-base estimate of the critical collapse functions. We also computed the NN-based estimates for critical collapse  functions under  $\omega=1.362$  and $\omega=1.003$ corresponding to estimates under  $\alpha$ and $\beta$ solutions. We treated these fixed NN-based estimates as the true forms of the critical collapse function under two different solutions to the equations of motion.  
We then applied the LOOCV technique to stack the NN candidates in estimating the critical functions when we used 100 random observations from the true NN-based estimates as the taring sets for the LOOCV-based NN methods. On the other side, we also obtained the 95\% Bayesian credible intervals for the critical collapse functions. To do that, we computed the 2.5 and 97.5 percentiles of the NN-based estimates, respectively, as the lower and upper bounds of the interval at each space-time point.     

\begin{figure}[H]
    \centering
    \includegraphics[scale=0.36]{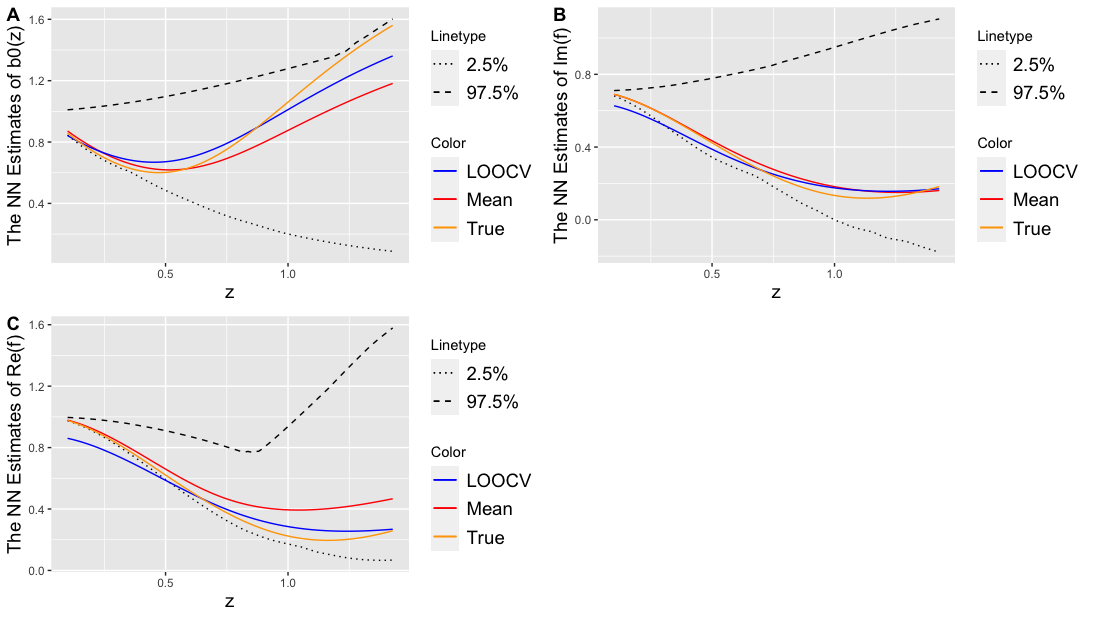}
    \caption{The posterior mean (red), LOOCV (blue) and the true, using $\omega=1.362$, (orange) NN-based estimates of the critical collapse functions $b_0(z)$ ({\bf A}), $\text{Im}(f(z))$ ({\bf B}) and  $\text{Re}(f(z))$ ({\bf C}) corresponding to the $\alpha$-solution scenario of the hyperbolic equations of motion. The dotted and dashed  lines show, respectively, the lower and upper bounds of the 95\% Bayesian credible intervals under Uniform prior.}
    \label{fig:unif_alpha_cv_3_1}
\end{figure}

\begin{figure}[H]
    \centering
    \includegraphics[scale=0.36]{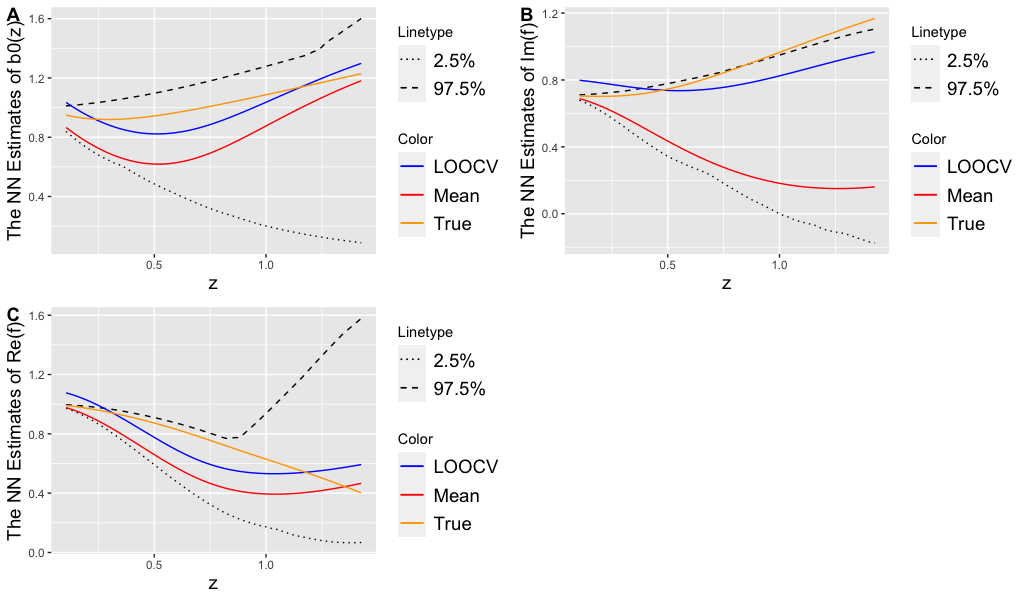}
    \caption{The posterior mean (red), LOOCV (blue) and the true, using $\omega=1.003$, (orange) NN-based estimates of the critical collapse functions $b_0(z)$ ({\bf A}), $\text{Im}(f(z))$ ({\bf B}) and  $\text{Re}(f(z))$ ({\bf C}) corresponding to the $\beta$-solution scenario of the hyperbolic equations of motion. The dotted and dashed  lines show, respectively, the lower and upper bounds of the 95\% Bayesian credible intervals under Uniform prior.}
    \label{fig:unif_beta_cv_3_1}
\end{figure}

\begin{figure}[H]
    \centering
    \includegraphics[scale=0.36]{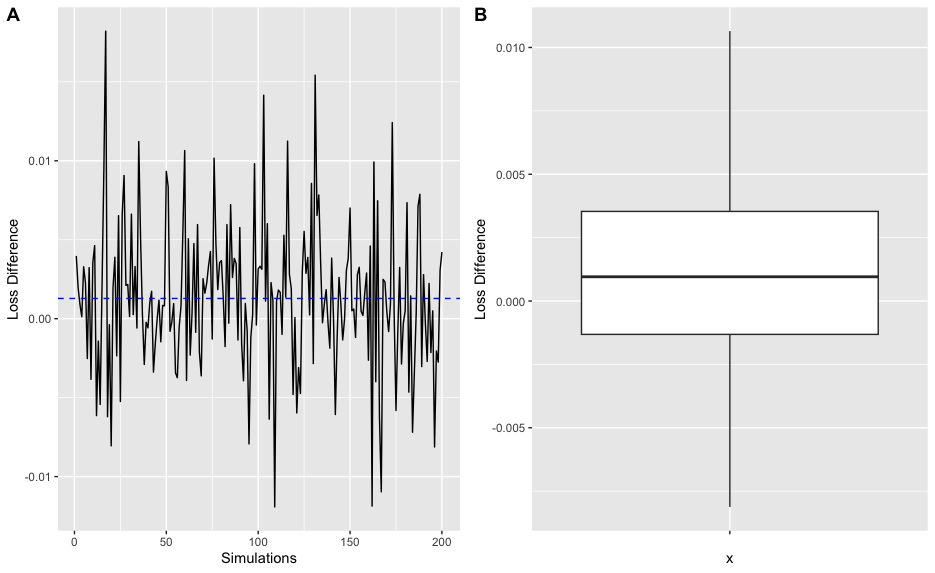}
    \caption{The trace- and box- plots of the differences between train and test loss values in the NN-based solvers using $L=200$ SMC samples from the posterior distribution of $\pi(\omega|{\bf y})$ when Uniform distribution was used as prior.}
    \label{fig:unif_loss_box}
\end{figure}

Figures \ref{fig:nor_alpha_3_1} and \ref{fig:unif_alpha_cv_3_1}
show the performance of the posterior mean NN-based estimates, the true NN-based estimates, the  LOOCV NN-based estimates as well as the 95\% credible intervals in estimating  the critical collapse functions corresponding to $\alpha$-solution under Gaussian and Uniform priors, respectively. 
Also Figures \ref{fig:nor_cv_beta_3_1} and \ref{fig:unif_beta_cv_3_1} show the results of their counterpart NN-based estimates corresponding to $\beta$-solution under Gaussian and Uniform prior distributions, respectively. 
Since the posterior mean NN-based estimates and 95\% credible intervals aggregate the $L=200$ posterior NN-based estimates  regardless of their corresponding population, these two methods remain robust and proposed the same estimates in predicting the critical collapse functions under both $\alpha$- and $\beta$- solutions to the equations of motion. 
Unlike posterior mean proposals, we recommend the LOOCV-based estimates if one is interested in aggregating the NN candidates to more accurately estimate the form and curvature of the critical collapse functions corresponding to specific solutions. 
We see that the LOOCV-based method, on average, more accurately estimates the critical functions for both $\alpha$ and $\beta$ solutions, because the method stacks the $L=200$ Bayesian NN candidates by developing the best  linear combination of the posterior NN candidates.   
\begin{figure}[H]
    \centering
    \includegraphics[scale=0.36]{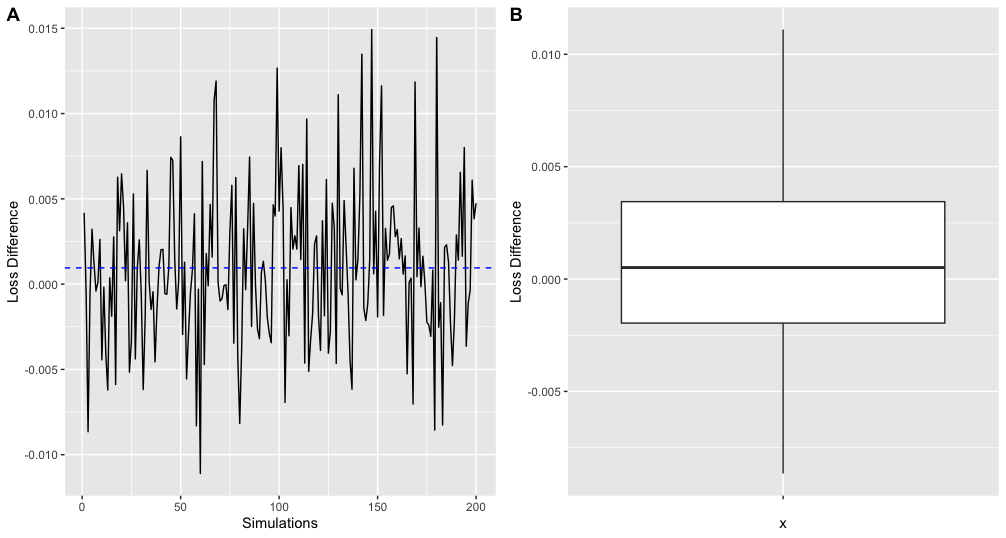}
    \caption{The trace- and box- plots of the differences between train and test loss values in the NN-based solvers using $L=200$ SMC samples from the posterior distribution of $\pi(\omega|{\bf y})$ when Gaussian distribution was used as prior.}
    \label{fig:nor_box}
\end{figure}
Last but not least, we investigated the convergence of the NN-based 
using SMC samples of $\omega$ under both Gaussian and Uniform prior distributions. 
To do that, we computed the difference between the training and test loss values in the last epoch of the NN-based solver for all the $L=200$ SMC samples $\omega^*_l,l=1,\ldots,1000$ from the posterior distributions. 
Figures \ref{fig:nor_box} and \ref{fig:unif_loss_box} report the trace and box plots of the loss differences when the Gaussian and Uniform distributions were, respectively, used as the prior distributions. 
From Figures \ref{fig:nor_box} and \ref{fig:unif_loss_box}, we observe that loss differences are almost less then $1.5\times10^{-2}$, fluctuating on average around zero for all the $L=200$ realizations.
This confirms the convergence of the NN-based solvers using SMC samples  after 1000 epochs even in  the hyperbolic equations of motion of four dimensions where the posterior distribution of the system parameter appear multi-modal with multiple high-density areas. 

Finally, it is important to highlight that if instead of taking continuous self-similarity, one only assumes discrete self-similarity, then the discrete scale transformation is compensated just by an element of $SL(2, Z)$ in the set $SL(2, R)$ transformation. It would also be very interesting to discover the fact that how all the critical exponents would depend on the modular transformations as well.

\section{Conclusions}\label{sec:conclusions}

This paper proposes a new formalism based on Sequential Monte Carlo (SMC) and artificial neural networks (NNs) to model the hyperbolic class of the spherical gravitational self-similar solutions in four dimensions. Due to the nature of highly non-linear ordinary differential equations for the axion-dilaton configurations,  in the literature, researchers typically have to employ various constraints as well as different numerical approximation methods to keep track of equations. For instance, in hyperbolic equations of motion, \cite{ours} had to apply the constraints, including the finiteness of $f''(z)$ as $z\rightarrow z_+$ and the vanishing of the divergent part of $f''(z)$ which generates a complex-valued constraint at $z_+$. They also employed numerical grid search discrete optimization methods on the extended coordinates of the equations to eliminate the spurious roots and estimate the equations' parameters and the self-similar black hole solutions.
Due to the sophisticated form of the equations and the vital role of the parameters, researchers usually have to overlook the measurement errors imposed in exploring the parameters through various numerical methods.

Here, we propose a new method to incorporate the measurement errors, involved in parameter estimation, into our statistical models in exploring the solutions to the equations of motion. Recently Hatefi et al. \cite{Hatefi:2023sgr} applied the Hamiltonian Monte Carlo method to find solutions to the equations of motion in the elliptic class of four dimension in a Bayesian framework. Unlike \cite{Hatefi:2023sgr}, the hyperbolic equations of motion in  four dimensions suffer from multiple solutions where the critical collapse functions have overlap domains under these solutions. To deal with this challenge,  for the first time in the literature on the axion-dilaton system, we proposed the SMC approach to derive the posterior distribution of the parameters.
The posterior distribution reveals all the possible solutions in estimating the parameter of the equations of motion. It is also important to highlight that the posterior distribution confirms the deterministic $\alpha$ and $\beta$ solutions found in the literature for the hyperbolic class in four dimensions. 
Unlike methods in the literature, in this paper, we proposed the $l_2$ penalized Leave-One-Out Cross-validation (LOOCV) to optimally combine the Bayesian NNs candidates. 
The approach enables us to determine the optimum weights while dealing with the co-linearity issue in the NN-based estimates and better predict the critical functions corresponding to multiple solutions of the hyperbolic equations of motion.

Using SMC samples from the posterior distribution, we then developed NN estimates based on the posterior mean and LOOCV 
as well as the 95\% credible intervals in estimating  the critical collapse functions corresponding to multiple solutions of the equations of motion. 
Because the posterior mean NN-based estimates and 95\% credible intervals aggregate all the posterior NN candidates   regardless of the multiple solutions of the system, these two methods remain robust and proposed the same estimates in predicting the critical collapse functions under both $\alpha$- and $\beta$- solutions.
Unlike the posterior mean proposal, we recommend the LOOCV-based estimates if one is interested in the optimum linear combination of the posterior NN candidates  to improve the estimating  of the critical collapse functions corresponding to established $\alpha$- and $\beta$-solutins in hyperbloic equations of motions in four dimensions.

Now if we compare the Bayesian method with the realistic  NN approach, then one gets to know that the developed Bayesian credible intervals actually contain the definite estimate as one possible candidate in the estimation of the critical collapse functions. Indeed, Unlike the estimation of \cite{Hatefi:2023vma}, the Bayesian approach remains concrete against measurement errors in estimating $\omega$ due to the fact that all these Bayesian estimations have already had all the possibilities of the parameter for the domain of the posterior distribution.  From a physical point of view, our results clarify that the universality of the Choptuik phenomena ~\cite{MA} is not satisfied. There may exist some universal behaviour which might be hidden in combining the critical exponents and other parameters of the theory. Nevertheless, our efforts provide some clear evidence that one cannot expect to transfer the standard expectations of Statistical Mechanics to the critical gravitational phenomena.

\section*{Acknowledgments}

 We would like to thank the editor and the anonymous referee for their valuable comments that improved the quality of the paper. E. Hatefi would like to especially thank Philip Siegmann for various discussions and supports. He is also grateful to R. López-Sastre for discussions. E. Hatefi also acknowledges A. Kuntz,  E. Hirschmann, L. Alvarez-Gaume, K. Narain, A. Sagnotti for valuable communications and supports. Parts of the work of E. Hatefi have been done during E. Hatefi's visit at Scuola Normale Superiore (SNS) in Pisa and he acknowledges SNS theory group.  E. Hatefi is also supported by the María Zambrano Grant of the Ministry of Universities of Spain. Armin Hatefi acknowledges the support from the Natural Sciences and Engineering Research Council of Canada (NSERC).


\section{Appendix}



\end{document}